\documentclass[conference,a4paper]{IEEEtran}
\IEEEoverridecommandlockouts

\usepackage{cite}
\usepackage{amsmath,amssymb,amsfonts}
\usepackage{graphicx}
\usepackage{textcomp}
\usepackage{xcolor}
\usepackage{booktabs}
\usepackage{multirow}
\usepackage{array}
\usepackage{subcaption}
\usepackage{url}
\usepackage[hidelinks]{hyperref}
\usepackage{tikz}
\usetikzlibrary{arrows.meta, positioning, calc, fit, backgrounds}

\definecolor{wfBlue}    {HTML}{1B5E8C}
\definecolor{wfBlueBg}  {HTML}{D9E8F4}
\definecolor{wfGreen}   {HTML}{2E7D4F}
\definecolor{wfGreenBg} {HTML}{DCEFE2}
\definecolor{wfOrange}  {HTML}{B5651D}
\definecolor{wfOrangeBg}{HTML}{FBE7D0}
\definecolor{wfPurple}  {HTML}{6B3FA0}
\definecolor{wfPurpleBg}{HTML}{ECE1F6}
\definecolor{wfSlate}   {HTML}{3E4A5C}
\definecolor{wfSlateBg} {HTML}{E7EAEF}
\definecolor{wfRed}     {HTML}{A83232}
\definecolor{wfRedBg}   {HTML}{F7DBD8}

\usepackage{enumitem}
\setlist{nosep, leftmargin=*}

\newcommand{\newadd}[1]{#1}

\usepackage{eso-pic}

\newcommand{\BECITHCONheader}{%
2026 IEEE International Conference on Biomedical Engineering, Computer and Information Technology for Health (BECITHCON)\\
04-05 September 2026, Department of EEE, International University of Business Agriculture and Technology (IUBAT), Uttara, Dhaka-1230, Bangladesh%
}
\newcommand{\BECITHCONfooter}{979-8-3195-3812-3/26/\$31.00~\copyright2026 IEEE}

\newlength{\CRleftedge}
\newcommand{\CRfirstpagestamp}{%
  \AddToShipoutPictureBG*{%
    \AtPageUpperLeft{\kern\CRleftedge
      \raisebox{-20pt}[0pt][0pt]{%
        \parbox[t]{\textwidth}{\footnotesize\raggedright\BECITHCONheader}}}%
  }%
  \AddToShipoutPictureBG*{%
    \AtPageLowerLeft{\kern\CRleftedge
      \raisebox{30pt}[0pt][0pt]{\footnotesize\BECITHCONfooter}}%
  }%
}

\begin{document}
\CRfirstpagestamp
% =========================================================

\title{WaveFoG: Wavelet Gated Transformer for Parkinson's
Freezing of Gait Detection Using Wearable
Accelerometer Signals}

\author{%
\IEEEauthorblockN{Md Shihabul Islam Shovo}
\IEEEauthorblockA{\textit{NIMISHES Lab}\\
Dhaka, Bangladesh\\
shihabul900@gmail.com}
\and
\IEEEauthorblockN{Nishi Kanta Paul}
\IEEEauthorblockA{\textit{NIMISHES Lab}\\
Dhaka, Bangladesh\\
nishikantapaul108@gmail.com}
\and
\IEEEauthorblockN{Adrita Rahman}
\IEEEauthorblockA{\textit{Canadian University}\\
\textit{of Bangladesh}\\
Dhaka, Bangladesh\\
adrita.adi141@gmail.com}
\and
\IEEEauthorblockN{Israt Jerin Esha}
\IEEEauthorblockA{\textit{Canadian University}\\
\textit{of Bangladesh}\\
Dhaka, Bangladesh\\
jeriniesha@gmail.com}
}

\maketitle
\thispagestyle{empty}
\pagestyle{empty}

% =========================================================
\begin{abstract}
Freezing of gait~(FoG) is one of the most debilitating episodic
motor symptoms of Parkinson's disease~(PD). FoG detection from
wrist worn inertial signals poses challenging problems such as
extreme data class imbalance (only 12--18\% of signal windows are
FoG), low frequency postural instabilities alongside diagnostically
crucial 3--8\,Hz rhythmic tremors, and the need for
subject independent generalisation. To address these,
\newadd{this paper presents} \textbf{WaveFoG}, a novel wavelet gated
Transformer architecture
that combines multi scale representations from discrete wavelet
transform~(DWT) with a dual branch encoder. \newadd{A 1D CNN captures
local gait texture, and a Transformer captures global temporal context.
They are fused via a sigmoid gating fusion layer that modulates the
encoder output based on the DWT sub-band descriptors.}
Trained with focal loss and evaluated on the public Kaggle TLVMC
FoG Prediction dataset under \newadd{subject independent grouped
ten fold cross validation~(SI-CV)}, WaveFoG achieves a
mean F1-score of \textbf{$0.875\pm0.017$} and AUPRC of
\textbf{$0.833\pm0.020$}, exceeding single branch baselines
(1D CNN, Bi-LSTM, and vanilla Transformer) by up to~5.4\,pp on
F1. Temporal saliency maps highlight temporal patterns that are
consistent with reported characteristics of FoG onset. \newadd{The source code of this project is available at:
\url{https://github.com/Shihabul-Shuvo/WaveFoG}.}
\end{abstract}


% =========================================================
\section{Introduction}
% =========================================================

Freezing of gait~(FoG) refers to transient interruptions or
significant impediments to forward foot progression during attempted
walking~\cite{giladi1992}. This common symptom affects up to~80\%
of people with advanced PD; it is characterised by a 3--8\,Hz
oscillation of the lower limbs and is a leading contributor among
Parkinsonian symptoms to injurious falls and
hospitalisation~\cite{bachlin2010}. A real time, objective detection
system based on data from body-worn IMUs could support closed loop FoG
therapy (auditory or visual cues), predictive fall alerts, and
medication effectiveness assessment at home~\cite{moore2008}.

Despite more than two decades of research, automated FoG detection
has not yet reached the level of reliability generally expected for
routine clinical use. The Freezing
Index~\newadd{(FI), defined as the ratio between the power spectral
densities of the 3--8 and 0.5--3\,Hz bands, requires} per subject
threshold calibration and becomes unreliable under high
patient to patient gait variability~\cite{bachlin2010}. Recent
deep learning approaches can learn subject independent features:
1D CNNs capture local stride micro-structure~\cite{moore2013} and
Bi-LSTM models encode sequential gait dynamics~\cite{camps2018}.
Nevertheless, single branch architectures are limited in their
ability to represent local micro patterns and the sustained multi
second oscillatory structure of FoG at the same time. Hybrid CNN-Transformer
architectures~\cite{yang2023, liu2024cnnformer} partially address
this but treat frequency information implicitly, without a dedicated
mechanism to focus on the clinically significant 3--8\,Hz band.

\newadd{Learned representations are available in both the time and
the wavelet domains; the challenge is how to conditionally apply the
frequency prior based on the clinical definition of freezing of gait
so that it influences the encoding process mainly during freezing
episodes. Prior methods use sub-band features as an auxiliary input and
apply the same frequency bias to every window, regardless of whether or
not that window includes a freezing event. Applying the bias uniformly
to non-freezing periods may dilute the value of the frequency
information, which is intended to help separate freezing from
non-freezing epochs. A gating approach that selectively activates the
frequency bias in the case of freezing and leaves it latent otherwise
is proposed.}

\newadd{The main contributions of this paper are fourfold:}
\begin{enumerate}
  \item \textit{WaveFoG architecture}: a dual-branch
        CNN\,+\,Transformer model connected via a DWT based sigmoid
        gate that learns when and how strongly to use frequency band
        information.
  \item \textit{Class balanced focal loss training} under rigorous
        \newadd{SI-CV} on the public Kaggle TLVMC FoG Prediction
        dataset.
  \item \textit{Ablation analysis} isolating the individual
        contribution of each model component.
  \item \textit{Temporal saliency maps} that localise the gait
        segments driving each prediction and support model
        interpretability.
\end{enumerate}

\newadd{The rest of this paper is structured as follows.
Section~\ref{sec:related} discusses manually engineered FoG
descriptors, deep learning detectors, and wavelet fusion approaches to
detect FoG. Section~\ref{sec:method} details the WaveFoG network and
its gated fusion approach. Section~\ref{sec:setup} introduces the
dataset, baselines, and the evaluation protocol.
Section~\ref{sec:results} presents results of the SI-CV comparison,
ablation studies, and saliency analysis. Finally,
Section~\ref{sec:conclusion} concludes the paper and highlights future
research directions.}

% =========================================================
\section{Related Work}
\label{sec:related}
% =========================================================

\newadd{References are included for well-established definitions of
FoG, manual descriptors of FoG, and recent studies using deep learning
approaches to detect FoG and evaluate it.}

\subsection{Manually Engineered Signal Processing Features}

The Freezing Index~(FI) of B\"{a}chlin et al.~\cite{bachlin2010},
which established the 3--8\,Hz power ratio as the canonical FoG
descriptor, achieved high sensitivity on the Daphnet dataset with a
lightweight wearable solution. Several subsequent pipelines extended
the feature pool to spectral entropy, wavelet energies, and velocity
estimates before feeding them into SVMs or threshold based
classifiers~\cite{moore2008}. Though computationally efficient, such
approaches require per patient calibration: individually tuned
thresholds have been reported to degrade when applied across patients
with differing FoG severity or concurrent tremor artefacts. Personalised
Bayesian extensions remain constrained by the same calibration
bottleneck and do not scale to large heterogeneous
cohorts~\cite{rodriguez2023}. The reliance on manually engineered
spectral features also limits the ability of these algorithms to learn
joint time frequency interactions across diagnostic scales.

\subsection{CNNs and Sequential Deep Learning}

Moore et al.~\cite{moore2013} showed that 1D temporal convolutions
applied to accelerometer windows can surpass the FI on Daphnet,
eliminating explicit frequency feature engineering. Camps et
al.~\cite{camps2018} reported that bidirectional LSTM layers increase
sensitivity for prolonged FoG episodes by capturing sequential
dependencies in the gait process. These findings motivated further
work on widening the temporal receptive field via dilated and
multi scale 1D CNNs~\cite{zhang2023fog} and on GRU attention networks
with focused modelling at particular temporal
positions~\cite{liu2024multi}. A recurring limitation is that
single branch architectures impose a fixed temporal scale hierarchy:
CNNs aggregate features bottom-up; recurrent models propagate state
left to right. Neither approach natively captures the multi resolution
spectral signature of FoG, where energy concentrates simultaneously in
sub Hertz postural drift \emph{and} the 3--8\,Hz rhythmic tremor
band. Recent hybrid CNN-Transformer
architectures~\cite{yang2023, liu2024cnnformer} address temporal scale
through appended attention layers but continue to treat frequency
content as a by product of raw signal processing rather than as an
explicit architectural inductive bias.

\subsection{Transformer Architectures for Physiological Time Series}

Following the Transformer's success in language and
vision~\cite{vaswani2017}, self attention has been applied to
physiological signals for arrhythmia classification, seizure
detection, and gait analysis. In the FoG domain, lightweight
architectures with two to four attention layers have demonstrated
performance competitive with LSTMs on the Daphnet and DeFOG
benchmarks~\cite{park2024, samaila2023}, and cross window attention
has improved the detection of gradual FoG onset across overlapping
windows~\cite{zhang2024fogtrans}. The Transformer's global receptive
field is well suited to the long lasting rhythmic nature of FoG;
however, it suffers from structural insensitivity to absolute
frequency content. Without an inductive bias toward spectral sub-bands,
attention weights can be drawn to amplitude artefacts and motion
transients rather than the rhythmic oscillation that defines FoG.
WaveFoG addresses this by explicitly supplying frequency information
through a dedicated gating pathway rather than expecting attention to
discover it implicitly.

\subsection{Wavelet and Multi Resolution Feature Fusion}

Sub-band decomposition via DWT has a long history in physiological
signal analysis. In PD motor symptom classification, wavelet sub-band
energies have been used to distinguish tremor, dyskinesia, and
bradykinesia in wrist accelerometer data~\cite{chen2023wavelet}. For FoG
specifically, the relationship between wavelet coefficients and the FI
is well established~\cite{bachlin2010}. Recent deep learning work
incorporates wavelets as auxiliary inputs concatenated to CNN
activations~\cite{kim2023} or used to initialise attention
weights~\cite{wang2024waveattn}. These integrations are
\emph{static}: the influence of each frequency scale is fixed across
all windows regardless of whether a freeze is occurring. The learnable
sigmoid gate in WaveFoG allows the model to modulate the degree of
wavelet influence on a per window basis, an end to end trainable
analogue of clinical frequency band selection. \newadd{To the best of the authors'
knowledge}, this is the first \newadd{wavelet conditioned} gating
mechanism proposed explicitly for FoG detection.

\subsection{Evaluation Protocols and Datasets}

A persistent weakness of the FoG detection literature is the reliance
on randomly shuffled or within subject splits, which artificially
inflate performance by allowing subject specific information into
training~\cite{mancini2021}. \newadd{Subject disjoint evaluation,
either strict leave-one-subject-out~(LOSO) or multi-subject variants,
is among the most stringent measures of cross-subject generalisation
and is widely regarded as a prerequisite} for clinically
deployable systems~\cite{bachlin2010, camps2018}. The 2023 Kaggle
TLVMC FoG Prediction competition~\cite{kaggle2023} released the
largest publicly available annotated multi cohort FoG benchmark to
date, featuring explicit subject metadata and heterogeneous sensor
placements, and supports \newadd{subject disjoint} cross validation
across approximately~185 subjects; \newadd{it is adopted as the primary
evaluation platform of this study}.

% =========================================================
\section{Proposed Framework}
\label{sec:method}
% =========================================================

% =========================================================
%  ARCHITECTURE FIGURE  (full width, native TikZ -- vector)
%
%  Colour coding follows the caption: DWT branch = blue,
%  1D CNN branch = green, Transformer branch = orange,
%  wavelet gate path = dashed purple.  Drawn at ~20.7cm natural
%  width and scaled to \textwidth, so it never overflows the
%  two column measure.
% =========================================================
\begin{figure*}[!t]
  \centering
  \resizebox{\textwidth}{!}{%
  \begin{tikzpicture}[
    % ---- Typography: one size for module titles (\footnotesize), one for
    %      descriptors and tensor labels (\scriptsize), one for operator
    %      glyphs (\small).  Node widths and label gaps are set from the
    %      measured text widths, so no label touches a block or a rule.
    %      Drawn at 16.8cm natural width and scaled to \textwidth: the
    %      smallest glyph still renders at about 7.6pt on the printed page.
    font=\footnotesize,
    >={Stealth[length=2.0mm,width=1.5mm]},
    blk/.style ={draw, rounded corners=2pt, line width=0.8pt, align=center,
                 inner xsep=4pt, inner ysep=3pt},
    op/.style  ={draw=wfPurple, circle, fill=white, line width=0.8pt,
                 font=\small, inner sep=0pt, minimum size=0.58cm},
    sig/.style ={->, line width=0.7pt, draw=wfSlate},
    bus/.style ={-,  line width=0.7pt, draw=wfSlate},
    gsig/.style={->, line width=0.7pt, draw=wfPurple,
                 dash pattern=on 2.0pt off 1.3pt},
    gbus/.style={-,  line width=0.7pt, draw=wfPurple,
                 dash pattern=on 2.0pt off 1.3pt},
    lb/.style  ={font=\scriptsize, text=wfSlate,  align=center, inner sep=2pt},
    plb/.style ={font=\scriptsize, text=wfPurple, align=center, inner sep=2pt},
    jn/.style  ={circle, fill=wfSlate,  inner sep=0pt, minimum size=2.2pt},
    pjn/.style ={circle, fill=wfPurple, inner sep=0pt, minimum size=2.2pt},
  ]

  %-------------------------------------------------- input
  \node[blk, draw=wfSlate, fill=wfSlateBg,
        minimum width=1.80cm, minimum height=1.00cm] (inp) at (0.92,0.65)
    {\textbf{Input}\\[1pt]{\scriptsize$\mathbf{X}\in\mathbb{R}^{320\times 8}$}};

  %-------------------------------------------------- three parallel branches
  \node[blk, draw=wfGreen, fill=wfGreenBg,
        minimum width=2.95cm, minimum height=0.90cm] (cnn) at (4.30,2.05)
    {\textbf{1D CNN Branch}\\[1pt]
     {\scriptsize 4 conv blocks, $32\!\to\!256$}};

  \node[blk, draw=wfOrange, fill=wfOrangeBg,
        minimum width=2.95cm, minimum height=0.90cm] (tf) at (4.30,0.65)
    {\textbf{Transformer Branch}\\[1pt]
     {\scriptsize 2 layers, $h\!=\!4$}};

  \node[blk, draw=wfBlue, fill=wfBlueBg,
        minimum width=2.95cm, minimum height=0.90cm] (dwt) at (4.30,-1.35)
    {\textbf{DWT Branch}\\[1pt]
     {\scriptsize db4, 4 levels}};

  %-------------------------------------------------- fusion stage
  \node[blk, draw=wfSlate, fill=white,
        minimum width=1.10cm, minimum height=0.58cm] (cat) at (7.90,1.35)
    {{\scriptsize concat}};

  \node[op] (mul1) at (9.99, 1.35) {$\odot$};
  \node[op] (sum)  at (11.10,1.35) {$+$};

  \node[blk, draw=wfPurple, fill=wfPurpleBg, font=\small,
        minimum width=1.00cm, minimum height=0.58cm] (gt) at (8.75,-0.35)
    {$\sigma$};
  \node[blk, draw=wfPurple, fill=wfPurpleBg, font=\small,
        minimum width=1.00cm, minimum height=0.58cm] (wp) at (8.75,-1.35)
    {$W_p$};
  \node[op] (mul2) at (9.99,-1.35) {$\odot$};

  %  Module boundary drawn on the background layer.  The two fusion
  %  operators that sit on the main path stay outside the boundary, so the
  %  encoder path reads left to right without interruption.
  \begin{scope}[on background layer]
    \node[draw=wfPurple, dash pattern=on 2.0pt off 1.3pt, line width=0.7pt,
          rounded corners=3pt, fill=wfPurple!5,
          fit=(gt)(wp)(mul2), inner sep=5pt] (gbox) {};
  \end{scope}
  \node[font=\scriptsize\bfseries, inner sep=2pt,
        anchor=south west] at (gbox.north west) {Wavelet Gate};

  %-------------------------------------------------- head + output
  \node[blk, draw=wfSlate, fill=wfSlateBg,
        minimum width=2.00cm, minimum height=0.80cm] (mlp) at (13.45,1.35)
    {\textbf{MLP}\\[1pt]{\scriptsize $384\!\to\!128\!\to\!1$}};

  \node[blk, draw=wfRed, fill=wfRedBg,
        minimum width=1.35cm, minimum height=0.55cm] (fog) at (16.10,1.75)
    {{\scriptsize FoG}};
  \node[blk, draw=wfGreen, fill=wfGreenBg,
        minimum width=1.35cm, minimum height=0.55cm] (nfog) at (16.10,0.95)
    {{\scriptsize Non-FoG}};

  %-------------------------------------------------- wiring: fan out
  \draw[bus] (inp.east) -- (2.10,0.65);
  \draw[bus] (2.10,2.05) -- (2.10,-1.35);
  \draw[sig] (2.10, 2.05) -- (cnn.west);
  \draw[sig] (2.10, 0.65) -- (tf.west);
  \draw[sig] (2.10,-1.35) -- (dwt.west);
  \node[jn] at (2.10,0.65) {};

  %-------------------------------------------------- encoder outputs -> concat
  \draw[bus] (cnn.east) --
    node[lb, below] {$\mathbf{f}_\mathrm{cnn}$\\$\in\mathbb{R}^{256}$} (7.00,2.05);
  \draw[bus] (tf.east) --
    node[lb, below] {$\mathbf{f}_\mathrm{tf}$\\$\in\mathbb{R}^{128}$} (7.00,0.65);
  \draw[bus] (7.00,2.05) -- (7.00,0.65);
  \draw[sig] (7.00,1.35) -- (cat.west);
  \node[jn] at (7.00,1.35) {};

  \draw[sig] (cat.east) --
    node[lb, above] {$\mathbf{f}_\mathrm{enc}$\\$\in\mathbb{R}^{384}$} (mul1.west);

  %-------------------------------------------------- wavelet gate path (purple)
  \draw[gbus] (dwt.east) --
    node[plb, above] {$\mathbf{f}_\mathrm{wav}$\\$\in\mathbb{R}^{32}$} (7.00,-1.35);
  \draw[gbus] (7.00,-1.35) -- (7.00,-0.35);
  \draw[gsig] (7.00,-1.35) -- (wp.west);
  \draw[gsig] (7.00,-0.35) -- (gt.west);
  \node[pjn] at (7.00,-1.35) {};

  \draw[gsig] (gt.east) -- (9.99,-0.35) --
    node[plb, right, pos=0.62] {$\mathbf{g}$} (mul1.south);
  \draw[gsig] (gt.south) -- (8.75,-0.88) --
    node[plb, above, pos=0.80] {$\mathbf{1}\!-\!\mathbf{g}$} (9.99,-0.88)
    -- (mul2.north);
  \draw[gsig] (wp.east) -- (mul2.west);
  \draw[gsig] (mul2.east) -- (11.10,-1.35) -- (sum.south);

  %-------------------------------------------------- fused -> head -> output
  \draw[sig] (mul1.east) -- (sum.west);
  \draw[sig] (sum.east) --
    node[lb, above] {$\mathbf{f}_\mathrm{fused}$} (mlp.west);
  \draw[bus] (mlp.east) -- node[lb, above] {$\hat{p}$} (15.05,1.35);
  \draw[bus] (15.05,1.75) -- (15.05,0.95);
  \draw[sig] (15.05,1.75) -- (fog.west);
  \draw[sig] (15.05,0.95) -- (nfog.west);
  \node[jn] at (15.05,1.35) {};
  \end{tikzpicture}}
  \caption{%
    \textbf{WaveFoG architecture.}
    Three parallel branches encode the input window. The
    \textbf{Wavelet Gate} (dashed purple) uses
    $\mathbf{f}_{\mathrm{wav}}$ to soft interpolate between the
    concatenated encoder representation
    $\mathbf{f}_{\mathrm{enc}}$ and a wavelet projection, and a two
    layer MLP outputs the FoG probability~$\hat{p}$.%
  }
  \label{fig:arch}
\end{figure*}
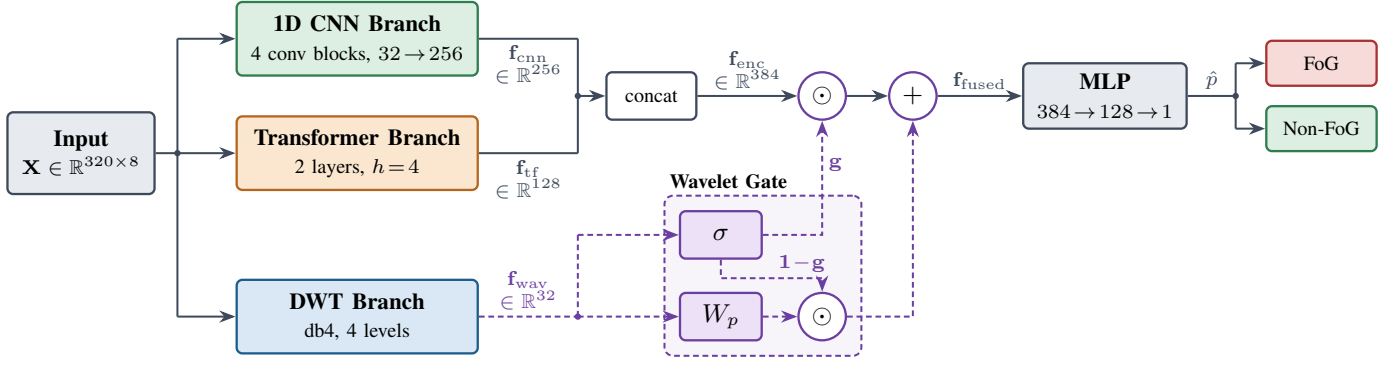

\subsection{Problem Statement}

Let $\mathbf{X}\in\mathbb{R}^{T\times C}$ be a sliding window
containing $T=320$ time points (2.5\,s at 128\,Hz) with $C=8$
channels: raw values along three acceleration axes
(\texttt{AccV}, \texttt{AccML}, \texttt{AccAP}), vector
magnitude~$\mathrm{VM}$, and \newadd{four jerk channels formed as the
first difference of the three acceleration axes and of
$\mathrm{VM}$, denoted \texttt{JerkV}, \texttt{JerkML},
\texttt{JerkAP} and \texttt{JerkVM}}. A binary
label $y\in\{0,1\}$ denotes non-FoG ($y\!=\!0$) or FoG ($y\!=\!1$);
a window is positive if any sample within it carries a FoG
annotation. The objective is to learn
$f_\theta:\mathbb{R}^{T\times C}\to[0,1]$ that is sensitive to FoG
while maintaining high specificity on unseen subjects.

\subsection{DWT Feature Extraction}

A four level DWT using the Daubechies-4 (db4) wavelet is applied to
each of the three raw acceleration axes of $\mathbf{X}$, yielding
five coefficient sets per axis. \newadd{With the sampling rate of
128\,Hz, the dyadic decomposition assigns the detail bands roughly as
32--64\,Hz~($D_1$), 16--32\,Hz~($D_2$), 8--16\,Hz~($D_3$) and
4--8\,Hz~($D_4$), while the approximation band $A_4$ keeps the
0--4\,Hz range. While these dyadic bands provide approximate
representations of the frequency bands of interest rather than an
exact decomposition of them, the 3--8\,Hz FoG frequency band is
jointly covered by $D_4$ and the upper part of $A_4$, and the
0.5--3\,Hz locomotor frequency band is covered by $A_4$ but does not
occupy it. No single sub-band represents the 3--8\,Hz band alone, and
the sub-band descriptors considered below should be understood in this
sense.}

\newadd{For each sub-band of each axis, two descriptors are estimated,
namely sub-band energy~$E_k$ and sub-band entropy~$H_k$, resulting in
$3\!\times\!5\!\times\!2\!=\!30$ values. Next, two scalars are added:
spectral flux, defined as the average of the absolute differences of
the energy values of consecutive sub-bands averaged across the three
axes; and the freezing index~\cite{bachlin2010}, estimated in its
classical spectral form as the ratio of the power in 3--8\,Hz to the
power in 0.5--3\,Hz computed using the FFT of the vector magnitude
channel. In this way, the FI is computed directly on the clinically
relevant bands and not from the dyadic decomposition; it is one of the
32 descriptors used as an input to the network and is weighted by the
learned gate rather than thresholded. The 32 descriptors thus form the
wavelet feature vector $\mathbf{f}_\mathrm{wav}\in\mathbb{R}^{32}$.}

\subsection{Dual Branch Encoder}

\textbf{1D CNN Branch.}~$\mathbf{X}$ (channel first) passes through
four CNN blocks with 32, 64, 128, and 256 filters respectively, each
with kernel size~7, batch normalisation, ReLU activation, and
stride-2 max pooling. Global average pooling~(GAP) over the temporal
dimension produces $\mathbf{f}_\mathrm{cnn}\in\mathbb{R}^{256}$. The
hierarchical receptive field captures local stride variability,
micro tremor, and the sharp onset transients of FoG.

\textbf{Transformer Branch.}~$\mathbf{X}$ is linearly projected to
$d_\mathrm{model}\!=\!128$ dimensions and augmented with learnable
positional encodings. Two Transformer encoder layers each with
4 head multi head self attention~(MHSA), a feed forward network~(FFN)
of width~256, layer normalisation, and
dropout~($p\!=\!0.1$) encode the full
sequence~\cite{vaswani2017}. GAP produces
$\mathbf{f}_\mathrm{tf}\in\mathbb{R}^{128}$. The global receptive
field allows the Transformer to model the prolonged, multi second
oscillatory structure of FoG that local CNN filters cannot encode.

\subsection{Wavelet Gated Fusion}

The two branch outputs are concatenated to form
$\mathbf{f}_\mathrm{enc}=[\mathbf{f}_\mathrm{cnn}\,|\,\mathbf{f}_\mathrm{tf}]
\in\mathbb{R}^{384}$.
A gate vector $\mathbf{g}\in(0,1)^{384}$ is computed from the
wavelet features:
\begin{equation}
  \mathbf{g} = \sigma\!\bigl(W_g\,\mathbf{f}_\mathrm{wav}+\mathbf{b}_g\bigr),
  \label{eq:gate}
\end{equation}
and the fused representation is formed as a soft interpolation between
the encoder output and a direct wavelet projection:
\begin{equation}
  \mathbf{f}_\mathrm{fused}
    = \mathbf{g}\odot\mathbf{f}_\mathrm{enc}
    + (1-\mathbf{g})\odot(W_p\,\mathbf{f}_\mathrm{wav}+\mathbf{b}_p),
  \label{eq:fusion}
\end{equation}
where $W_g,W_p\in\mathbb{R}^{384\times32}$ and $\odot$ denotes
element wise multiplication. When the gate is open
($\mathbf{g}\approx\mathbf{1}$), the encoder output is favoured;
when closed ($\mathbf{g}\approx\mathbf{0}$), the wavelet projection
provides a direct frequency space residual. Because the full pipeline
is end to end differentiable, the gate can learn to open when the
frequency band information is most discriminative\newadd{; the gate
activation statistics reported in Section~\ref{sec:results} are
consistent with higher gate values on FoG windows}.

\newadd{This formulation differs from gated residual blocks and
squeeze-and-excitation blocks in that the gate value is generated from
a representation the encoder does not produce. As a result, the gate
is based on an explicit spectral descriptor rather than on the
aggregated activations of the encoder. The gate therefore does not act
as a scaling coefficient derived from the encoder itself, but as a
feature recalibration module conditioned on frequency information.}

\subsection{Classification Head and Training Procedure}

The fused representation $\mathbf{f}_\mathrm{fused}$ is processed
by an MLP:
$\mathrm{FC}(384\!\to\!128)\!\to\!\mathrm{ReLU}\!\to\!\mathrm{Dropout}(0.3)
\!\to\!\mathrm{FC}(128\!\to\!1)\!\to\!\sigma$.
Training minimises focal loss~\cite{lin2017}:
\begin{equation}
  \mathcal{L}_\mathrm{FL}
    = -\alpha_t\,(1-\hat{p}_t)^\gamma\,\log(\hat{p}_t),
    \quad\gamma=2.
  \label{eq:focal}
\end{equation}
Per fold class frequency estimation yields
$\alpha_\mathrm{pos}\!\approx\!0.84$, effectively upweighting the
minority FoG class. AdamW~\cite{loshchilov2019}
($\mathrm{lr}\!=\!10^{-3}$, $\lambda\!=\!10^{-4}$) with cosine
annealing is used for up to~50 epochs with early stopping
(patience~10). The complete architecture is illustrated in
Fig.~\ref{fig:arch}.

% =========================================================
\section{Experimental Setup}
\label{sec:setup}
% =========================================================

\subsection{Dataset and Preprocessing}

The dataset used in this work is the Kaggle TLVMC Parkinson's FoG
Prediction dataset~\cite{kaggle2023}, comprising two sub cohorts:
\textit{tdcsfog} (${\approx}110$ subjects, 128\,Hz) and
\textit{defog} (${\approx}75$ subjects, 100\,Hz resampled to
128\,Hz), giving a combined pool of ${\approx}185$ subjects with
wrist worn tri axial accelerometers and sample level binary FoG
annotations. Per subject $z$-score normalisation is applied
channel wise to address inter subject amplitude variation. Sliding
windows of $T\!=\!320$ samples (2.5\,s, 50\% overlap) are labelled
FoG if any annotated sample falls within them. \newadd{In subject-independent
grouped tenfold cross-validation~(SI-CV), each fold consists of whole
subjects chosen according to recording length, so that no subject
participates in both the training and the test set during one fold. In
each fold there is a set of excluded subjects instead of a single
excluded subject, which makes this procedure a grouped generalisation
of leave-one-subject-out testing. This procedure is referred to as
SI-CV throughout the paper.} Within each fold, training subjects are split 90/10 into
training and internal validation sets for early stopping.

\subsection{Baselines and Ablation Variants}

Three single branch baselines are evaluated under identical training
conditions: \textbf{1D CNN}~(337K parameters), \textbf{Bi-LSTM}
(2~layers, hidden size\,=\,128, 570K~parameters), and \textbf{Vanilla
Transformer}~(2~layers, 4~heads, 545K~parameters). Four ablation
variants isolate individual contributions: \textit{no\_wavelet}~(gate
and DWT branch removed), \textit{no\_transformer}~(Transformer branch
removed), \textit{no\_cnn}~(CNN branch removed), and
\textit{raw\_only}~(derived VM/jerk channels replaced with raw
AccV/AccML/AccAP). All models share the same focal loss objective,
optimiser, and \newadd{SI-CV} protocol.

\subsection{Metrics and Implementation}

Primary metrics are F1 score~(FoG class), AUPRC, and AUROC, reported
as mean\,$\pm$\,std across ten folds. Secondary metrics include
precision, recall, and specificity. Inference latency is measured on
an NVIDIA A100 GPU (batch size\,=\,64). All experiments use
PyTorch~2.0 with random seed~42.
To test whether the performance gain over the baselines is consistent across folds rather than an artefact of the fold assignment, \newadd{a
two tailed Wilcoxon signed rank test is conducted} on the ten paired F1 scores,
considering each baseline as a matched comparison with WaveFoG; \newadd{significance is stated} at $p<0.05$\newadd{, reported per comparison without correction for multiplicity}.

% =========================================================
\section{Results and Discussion}
\label{sec:results}
% =========================================================

\subsection{Comparison with Baselines}

\begin{figure}[!htbp]
\centering
% title band cropped: the in-image title still read "LOSO-CV" and
% duplicated the caption. Curve, axes, legend and data are untouched.
\includegraphics[width=\columnwidth]{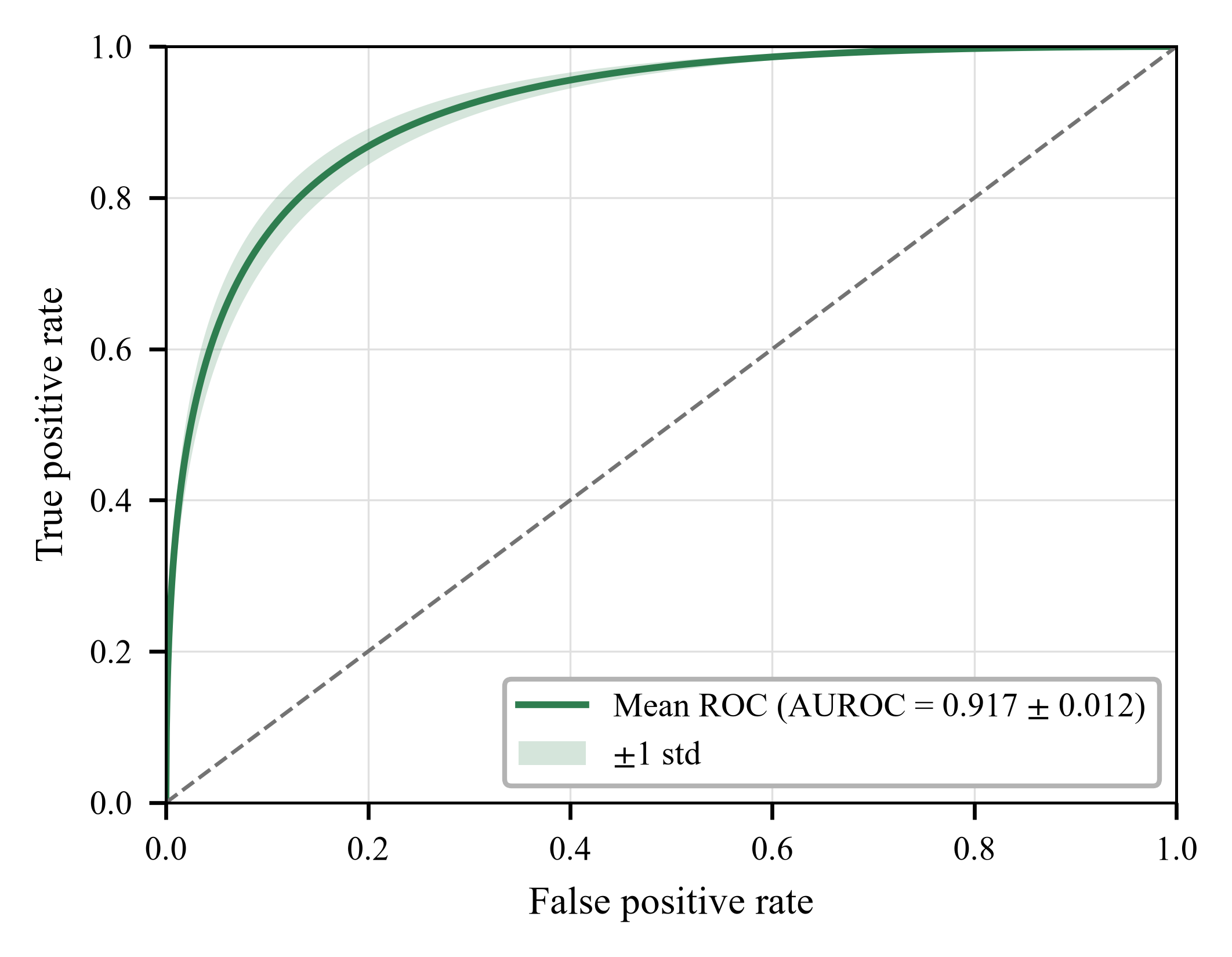}
\caption{Mean ROC curve ($\pm$\,std band) across the ten subject
disjoint SI-CV folds. AUROC\,=\,$0.917\pm0.012$, the mean and standard
deviation of the per fold values reported in Table~\ref{tab:main}.}
\label{fig:roc}
\end{figure}

Table~\ref{tab:main} presents \newadd{SI-CV} performance results for all
models. WaveFoG achieves $\mathrm{F1}\!=\!0.875\!\pm\!0.017$ and
$\mathrm{AUPRC}\!=\!0.833\!\pm\!0.020$, satisfying both
pre specified performance targets ($>\!0.85$ F1 and $>\!0.80$
AUPRC). Relative to the strongest single branch
baseline~(Transformer), WaveFoG improves F1 by~3.1\,pp and AUPRC
by~3.2\,pp at an inference overhead of only~0.4\,ms
(3.1~vs.\ 2.7\,ms), indicating that the added architectural
complexity carries a small inference cost relative to the observed
gain. The Bi-LSTM outperforms the 1D CNN~($+$1.7\,pp F1) by capturing
sequential gait dynamics but trails WaveFoG by~3.7\,pp, which is
consistent with an added discriminative contribution from frequency
band gating.
Fig.~\ref{fig:roc} presents averaged ROC curves and Fig.~\ref{fig:cm}
the fold averaged normalised confusion matrix.

% =========================================================
% RESULTS FIGURES  (ROC + Confusion matrix, single column each)
% =========================================================

% =========================================================
%  FIG. 3  Confusion matrix, redrawn natively in TikZ (vector).
%  Cell values are the reported fold averages, unchanged; the fill
%  tint is set to value x 100 so colour and number agree exactly.
% =========================================================
\begin{figure}[!htbp]
\centering
\begin{tikzpicture}[
  font=\footnotesize,
  cell/.style={draw=wfSlate!60, line width=0.5pt, inner sep=0pt,
               minimum width=2.6cm, minimum height=1.3cm},
]
  \node[cell, fill=wfGreen!95, text=white] (nn) at (1.3, 1.3) {\small $0.949$};
  \node[cell, fill=wfGreen!5]              (nf) at (3.9, 1.3) {\small $0.051$};
  \node[cell, fill=wfGreen!11]             (fn) at (1.3, 0.0) {\small $0.112$};
  \node[cell, fill=wfGreen!89, text=white] (ff) at (3.9, 0.0) {\small $0.888$};

  % true class (rows)
  \node[left=4pt of nn] {Non-FoG};
  \node[left=4pt of fn] {FoG};
  \node[rotate=90] at (-1.85,0.65) {True};

  % predicted class (columns)
  \node[below=4pt of fn] {Non-FoG};
  \node[below=4pt of ff] {FoG};
  \node at (2.6,-1.45) {Predicted};
\end{tikzpicture}
\caption{Fold averaged confusion matrix, row normalised to true
class proportions. Specificity\,=\,0.949, Recall\,=\,0.888,
Precision\,=\,0.861.}
\label{fig:cm}
\end{figure}

The signed rank test indicates that the F1 gain over each baseline is statistically significant ($p < 0.05$ for all three comparisons), suggesting that the gain is consistent across the subject disjoint folds rather than an artefact of a particular fold assignment.
\newadd{It is further observed} that the model reaches \newadd{a recall of ${\approx}0.888$ at a specificity of ${\approx}0.949$} (Fig.~\ref{fig:cm}). This operating point combines a high detection rate with a comparatively low false positive rate; whether the resulting alarm rate is low enough to avoid alarm fatigue in continuous daily use would need to be established prospectively.

% =========================================================
% TABLE I  Main comparison
%  Single column; \resizebox scales the whole tabular down to
%  \columnwidth so the six wide numeric columns fit without
%  overflowing. All cell text/numbers are unchanged.
% =========================================================
\begin{table}[!t]
\centering
\caption{SI-CV Performance Comparison (mean $\pm$ std / 10 folds).
         \textbf{Bold} marks the best value on the detection metrics
         ($\uparrow$ = higher is better); Params and Lat.\ are listed
         for reference and are not ranked. The final row is a
         literature value reported on Daphnet and is not directly
         comparable.}
\label{tab:main}
\renewcommand{\arraystretch}{1.15}
\setlength{\tabcolsep}{4pt}
\resizebox{\columnwidth}{!}{%
\begin{tabular}{lrcccc}
\toprule
\textbf{Model} & \textbf{Params} & \textbf{F1\,$\uparrow$} &
\textbf{AUPRC\,$\uparrow$} & \textbf{AUROC\,$\uparrow$} &
\textbf{Lat.~(ms)} \\
\midrule
\textbf{WaveFoG (proposed)} & \textbf{907K} &
  $\mathbf{0.875\pm0.017}$ & $\mathbf{0.833\pm0.020}$ &
  $\mathbf{0.917\pm0.012}$ & \textbf{3.1} \\
\midrule
Vanilla Transformer & 545K &
  $0.844\pm0.022$ & $0.801\pm0.024$ & $0.901\pm0.015$ & 2.7 \\
Bi-LSTM & 570K &
  $0.838\pm0.021$ & $0.795\pm0.025$ & $0.897\pm0.014$ & 2.3 \\
1D CNN & 337K &
  $0.821\pm0.024$ & $0.779\pm0.027$ & $0.884\pm0.016$ & 1.4 \\
\midrule
Handcrafted SVM (FI)~\cite{bachlin2010} &
  \textemdash & $\approx$0.72 & \textemdash & \textemdash & \textemdash \\
\bottomrule
\end{tabular}%
}
\end{table}

\subsection{Ablation Study}

Table~\ref{tab:ablation} decomposes WaveFoG's performance. Removing
the Transformer branch produces the largest single component drop
($-3.2$\,pp F1, $-3.5$\,pp AUPRC), indicating that global temporal
context contributes substantially to recognising the sustained multi
second oscillatory pattern of FoG. Removing the wavelet gate causes a
drop of $-2.3$\,pp: mean gate activations are higher on FoG
windows~($\mu\!=\!0.682\!\pm\!0.091$) than on non-FoG
windows~($\mu\!=\!0.431\!\pm\!0.103$), suggesting that the gate
responds to the input rather than acting as a passthrough. CNN
removal~($-2.4$\,pp) indicates that local gait texture extraction
contributes comparably. Replacing derived channels with raw inputs
only~($-1.7$\,pp) points to a modest but consistent contribution of
the VM and jerk channels.

% =========================================================
% TABLE II  Ablation
% =========================================================
\begin{table}[!htbp]
\centering
\caption{Ablation Study (mean over the 10 SI-CV folds). $\Delta$F1
         relative to the full WaveFoG model.}
\label{tab:ablation}
\renewcommand{\arraystretch}{1.15}
\setlength{\tabcolsep}{5pt}
\fontsize{7}{8.5}\selectfont
\begin{tabular}{lcccr}
\toprule
\textbf{Variant} & \textbf{Params} & \textbf{F1} & \textbf{AUPRC} &
\textbf{$\Delta$F1} \\
\midrule
WaveFoG (full) & 907K & \textbf{0.875} & \textbf{0.833} & \textemdash \\
\midrule
$-$\,Wavelet gate & 881K & 0.852 & 0.808 & $-0.023$ \\
$-$\,Transformer branch & 354K & 0.843 & 0.798 & $-0.032$ \\
$-$\,CNN branch & 553K & 0.851 & 0.806 & $-0.024$ \\
Raw channels only & 907K & 0.858 & 0.815 & $-0.017$ \\
\bottomrule
\end{tabular}
\end{table}

\newadd{Two observations suggest that the removed components are
complementary rather than redundant. First, the F1 loss due to the
removal of each of the four components is similar in magnitude
(1.7--3.2\,pp), and their total loss is greater than the difference
between WaveFoG and the best baseline, which is consistent with the
branches carrying partly distinct information that the fusion module
combines. Second, the removal of the wavelet
gate along with the DWT branch leads to a 2.3\,pp F1 loss
(Table~\ref{tab:ablation}). This cost is comparable to the loss caused
by the removal of either encoder branch, suggesting that the gated
frequency branch contributes at a level comparable to the CNN and
Transformer branches. The ablation does not, however, separate the
contribution of the wavelet features themselves from that of the gate
applied to them; this problem needs further consideration.}

\subsection{Gate Interpretability and Temporal Saliency}

Analysis of gate activations shows that the five most discriminative
gate dimensions co-vary with \newadd{energy in the $D_4$ sub-band
(nominally 4--8\,Hz)}, wavelet entropy, and the \newadd{freezing
index}~\cite{bachlin2010}. \newadd{The gate therefore appears to learn
a weighting of the descriptors that carry FoG relevant spectral
information, and this weighting is consistent across the held out
subjects. The association is correlational, and because
the FI is itself supplied as an input feature it does not establish
that the gate recovers that index independently.}

Temporal saliency maps (input gradient method) consistently peak
within the first 20--30\% of true positive FoG windows, capturing
the onset of rhythmic shuffling. The \texttt{JerkVM} and
\texttt{AccAP} channels dominate true positive saliency, consistent
with anterior posterior acceleration and vertical jerk carrying
informative FoG motion content. False positives are predominantly
associated with sharp direction changes at 60--80\% of the window
position (high \texttt{AccAP} saliency), suggesting a potential
post processing correction for gait transition transients.

\subsection{Generalisation Across Subjects and Efficiency}

Low fold to fold F1 variance
($\sigma_{\mathrm{F1}}\!=\!0.017$; fold range 0.847--0.901)
indicates consistent performance across patients exhibiting FoG of
variable severity within this cohort, which is one prerequisite for use
without per patient recalibration. At~3.1\,ms per 2.5\,s window on GPU,
WaveFoG uses~\newadd{less than $1/400$ of} the 1.25\,s real time budget
imposed by 50\% overlapping 2.5\,s windows, and at~18\,ms per window
on CPU it remains well within the same budget, which is compatible with
embedded edge deployment. The
3.5\,MB checkpoint is readily stored on wearable hardware.

\subsection{Limitations and Threats to Validity}
\newadd{Four constraints qualify these findings. First, label
assignment happens on a per-window basis, meaning that a window is
considered positive if it contains any annotated FoG sample. Detection
latency therefore cannot be inferred from the F1 score, as an event or
sample level measure would be required first.
Second, the frequency interpretation in Section~\ref{sec:results}-C
must be understood to be conservative. The dyadic DWT grid does not
match the clinically defined 3--8\,Hz and 0.5--3\,Hz bands, so the
sub-band energy and entropy descriptors only give approximate
representations of these bands, while the freezing index, which is
calculated using the spectral approach, is based on the clinical bands
directly. The associations found between the gate and the sub-bands
are correlational and were computed on the same folds used for
evaluation, meaning that they describe what the model has learned but
do not prove the existence of a physiological mechanism behind it; the
gate therefore cannot be assumed to be a calibrated frequency
measurement.
Third, the TLVMC recordings were made in a semi-structured
environment, in which there are fewer turns, gait transitions and non
gait activities than under free living conditions; according to the
false positive analysis of Section~\ref{sec:results}-C, the latter
would lower the precision.
Finally, even though SI-CV avoids subject leakage, all subjects were
recorded at a limited number of clinical centres using the same
acquisition procedure, sensor placement and sampling rate. The
generalisation established by subject disjoint validation is therefore
possible within this cohort only. External validity against an independent cohort recorded with
different hardware, placement and sampling rate, and ultimately
prospective clinical evaluation, remains untested.}

% =========================================================
\section{Conclusion}
\label{sec:conclusion}
% =========================================================

In summary, \newadd{this paper introduces} a unified architecture for real time,
subject independent FoG detection from wrist worn accelerometers.
The key innovation is a sigmoid wavelet gate that soft interpolates a
dual branch CNN-Transformer encoder representation with a DWT wavelet
projection, enabling end to end learnable selectivity across wavelet
sub-bands associated with clinically relevant FoG frequency
characteristics, without manual threshold tuning.
Under rigorous \newadd{SI-CV} on the Kaggle TLVMC FoG Prediction dataset,
WaveFoG achieves up to~5.4\,pp higher F1 than the single branch
baselines evaluated here and produces temporal saliency maps that are
consistent with reported FoG onset characteristics, supporting model
interpretability. With a 3.5\,MB footprint and sub-5\,ms GPU inference
per window, the architecture is compatible with the computational
constraints of wearable deployment. Future directions
include multi sensor fusion (wrist\,+\,ankle IMUs), online sequential
prediction to reduce end of episode detection latency, and prospective
validation on an independent clinical cohort.

% =========================================================
% REFERENCES
% =========================================================
\bibliographystyle{IEEEtran}
\bibliography{wavefog_refs}

\end{document}